\documentclass{svjour3}                     
\smartqed  
\usepackage{graphicx} 

\begin{document} 

\newcommand{\be}{\begin{equation}}
\newcommand{\ee}{\end{equation}}
\newcommand{\bea}{\begin{eqnarray}}
\newcommand{\eea}{\end{eqnarray}}

\newcommand{\re}{\mathop{\mathrm{Re}}}

\title{Non-exotic conformal structure of weak exotic singularities}

\author{Mariusz P. D\c{a}browski        \and 
Konrad Marosek
}

\institute{Mariusz P. D\c{a}browski \at 
Institute of Physics, University of Szczecin, Wielkopolska 15, 70-451 Szczecin, Poland \\
National Centre for Nuclear Research, Andrzeja So{\l}tana 7, 05-400 Otwock, Poland\\
 Copernicus Center for Interdisciplinary Studies, S{\l }awkowska 17, 31-016 Krak\'ow, Poland\\
 \email{Mariusz.Dabrowski@usz.edu.pl}
\and 
Konrad Marosek \at 
Chair of Physics, Maritime University, Wa{\l}y Chrobrego 1-2, 70-500 Szczecin, Poland \\
Institute of Physics, University of Szczecin, Wielkopolska 15, 70-451 Szczecin, Poland\\
\email{Konrad.Marosek@usz.edu.pl}
}






\date{Received: date / Accepted: date}

\maketitle

\begin{abstract}
We study the conformal structure of exotic (non-big-bang) singularity universes using the hybrid big-bang/exotic singularity/big-bang and big-rip/exotic singularity/big-rip models by investigating their appropriate Penrose diagrams. We show that the diagrams have the standard structure for the big-bang and big-rip and that exotic singularities appear just as the constant time hypersurfaces for the time of a singularity and because of their geodesic completeness are potentially transversable. We also comment on some applications and extensions of the Penrose diagram method in studying exotic singularities. 
\end{abstract}

\section{Introduction}
\label{intro}
One of the main obstacles in general relativity are the singularities which were described in the most general way by the notion of geodesic incompleteness \cite{HE}. The nature of singularities is, however, more sophisticated and various tools to study them were suggested. Among them the integral definitions of the weak and strong singularities given by Tipler \cite{tipler} and Kr\'olak \cite{krolak}. Their practical use was not very much explored until the discovery of dark energy \cite{supernovae} and in particular the phantom, which leads to a strong singularity -- a big-rip \cite{phantom} -- in that sense similar to a big-bang. Growing interest in various forms of dark energy uncovered other types of singularities -- most of them of a weak nature. The very paper of Barrow \cite{Barrow04} presented the sudden future singularity (SFS) of pressure (also called a big-brake and in fact being a subcase of an SFS \cite{big-brake}) which was given some observational studies \cite{obsSFS}. Many other studies of these singularities followed \cite{exotic}. 

More weak singularities were first investigated in Ref. \cite{nojiri} (finite scale factor singularity, big-separation) -- classified as types I-IV, and later appended in Refs. \cite{wsing,LRip,PRip} (w-singularity, little-rip, pseudo-rip). The full classification of the standard and exotic singularities in homogeneous and isotropic Friedmann universes was presented in Refs. \cite{APS2010,limits} (for the discussion of non-homogeneous models with exotic singularities see e.g. \cite{PRD05}). One of the issues is whether the weak exotic singularities can be transversable in the sense of geodesic parameter \cite{adam,LFJ2007,leonardo1,perivolaropoulos}. Recently, even the discussion of the transition through strong (big-bang) singularities was performed \cite{kamen17,brandenb18}. 

In this paper we investigate the conformal structure of the spacetimes with weak exotic singularities by using the method of Penrose diagrams. We follow the discussion of Ref. \cite{harada} for strong singularities such as the big-bang and the big-rip. 

\section{Conformal transformations and Penrose diagrams}

We use the Penrose diagram method and start with Friedmann $(k = 0)$ metric:
\be
ds^2=-c^2 dt^2 + a^2 \left( t \right) \left[ dr^2 + r^2 \left( d {\theta}^2 +{\sin}^2 \theta d{\phi}^2 \right) \right],
\label{FRW}
\ee
which after the application of the conformal time
\be
\eta= \int \frac{c dt}{a \left( t \right)}
\label{conftime}
\ee
can be transformed into
\bea
ds^2 &=& d \hat{s}^2 a^2 \left( \eta \right) \\
&=& a^2 \left( \eta \right)  \left[ -d{\eta}^2 +  dr^2 +  r^2 \left( d {\theta}^2 +{\sin}^2 \theta d{\phi}^2 \right) \right] , \nonumber
\label{transf}
\eea
where
\be
d{\hat{s}}^2= -d{\eta}^2 +  dr^2 +  r^2 \left( d {\theta}^2 +{\sin}^2 \theta d{\phi}^2 \right)
\label{Mink}
\ee
is the Minkowski metric. Using the following coordinate transformations ($ 0\le r \le \infty $)
\bea
t' &=& \arctan{(\eta + r)} + \arctan{(\eta - r)} ,
\label{MinkEt}\\
r' &=&  \arctan{(\eta + r)} - \arctan{(\eta - r)} .
\label{MinkEr}
\eea
one maps the Minkowski metric (\ref{Mink}) onto the Einstein static universe with the radius $r_E = \sin{r'}$ i.e.
\be
d\breve{s}^2=-dt'^2 + dr'^2 + \sin ^2 r' \left( d {\theta}^2 +{\sin}^2 \theta d{\phi}^2 \right) .
\label{ESU}
\ee
When the projection of a model is given, then one is able to draw the Penrose diagram \cite{HE}. 

\section{Conformal structure of weak exotic singularities}

\subsection{Hybrid big-bang/exotic singularity models}

\begin{figure}[htbp]
\includegraphics[width=8.3cm]{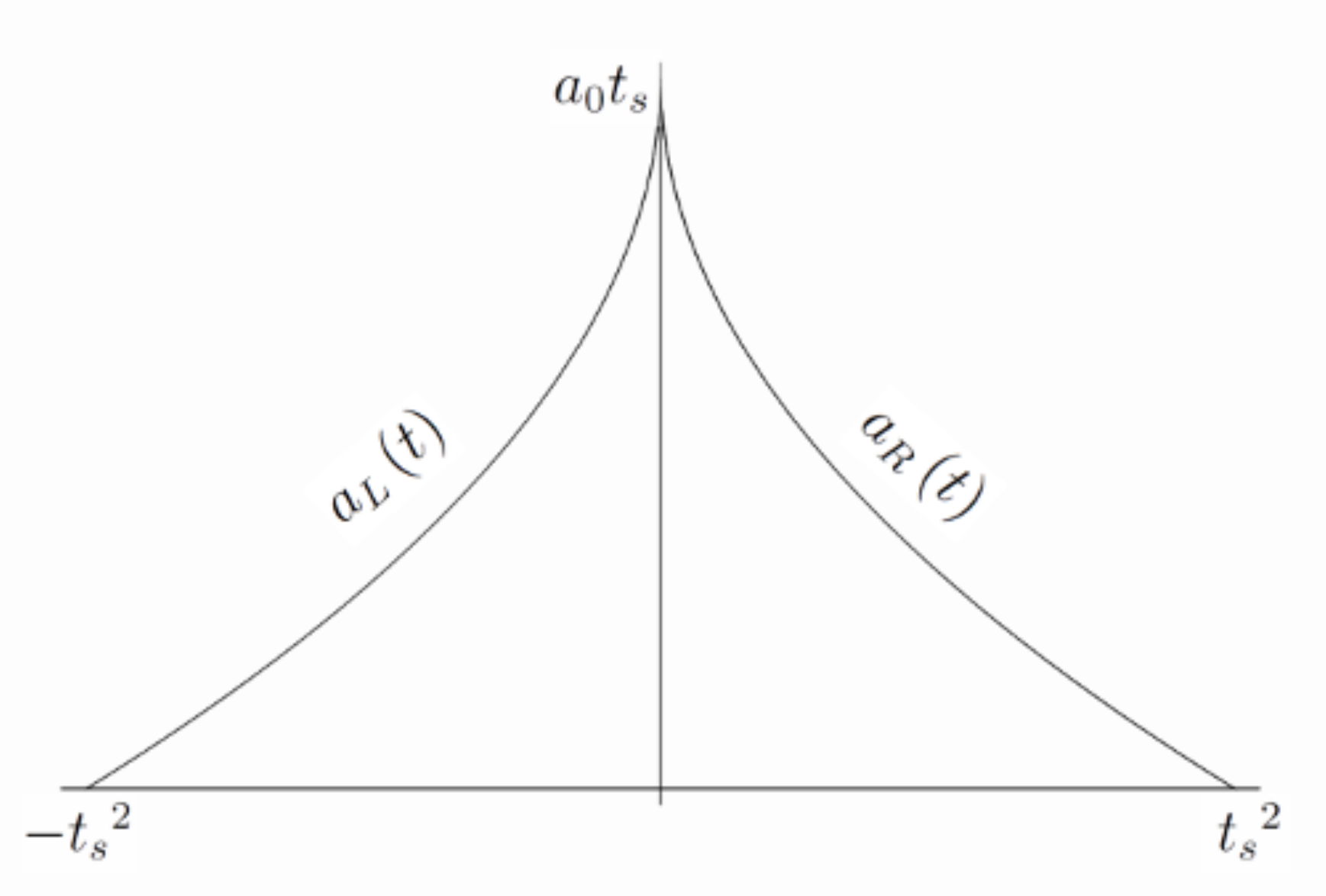}
\caption{The scale factor for the model (\ref{EgzoticScaleLR}) with $ n = 1/2 $. The evolution starts with a big-bang, reaches an exotic singularity, and finally ends at a big-crunch.}
\label{scalefactor}
\end{figure}

The following scenario for the universe evolution was suggested in Ref. \cite{PLB11}: it starts with a big-bang, reaches an exotic singularity, and then continues to a big-crunch. The scale factor is composed of the two branches (cf. Fig. \ref{scalefactor}) and reads as 
\be
a_{L,R} (t) = a_s \left[\delta + \left(1 \pm \frac{t}{t_s} \right)^m \left(1 - \delta \right) - \delta \left(\mp \frac{t}{t_s} \right)^n \right]
\label{aLRB}
\ee
with a big-bang $a_L(-t_s)=0$, a sudden future singularity $a_{L}(0)=a_{R}(0) = a_s$, and a big-crunch time, $a_R(t_s)=0$, and $a_s, \delta, m=$ const., $1<n<2$. A different form of the scale factor (\ref{aLRB}) was proposed in Ref. \cite{JCAP13}. After appropriate shift of the exotic singularity it can be written down as
\bea
\label{OurSF}
a_{L,R} \left( t \right) = a_0 {\left( \frac{\pm t}{t_s} + 1 \right)}^m \exp \left[ {\left(\frac{\mp t}{t_s} \right)}^n \right]~,
\eea
where the big-bang/big-crunch appears at $ t \to \mp t_s $, and an exotic singularity in $ t \to 0 $. By an appropriate choice of the parameter $n$ the scale factor (\ref{OurSF}) describes a sudden future singularity ($1<n<2$), a finite scale factor singularity ($0<n<1$), a big-separation ($2<n<3$), and a $w-$singularity ($3<n<4$) \cite{JCAP13}. 

The description of a transition from (\ref{FRW}) to (\ref{ESU}) with the scale factors (\ref{aLRB}) and (\ref{OurSF}) is impossible analytically. So following the approach of Ref. \cite{PLB11} in order to investigate the conformal structure of these models, we apply a simpler form of the scale factor which allows both an exotic singularity (depending on the value of the parameter $n$) and a standard big-bang singularity which reads as
\be
\label{EgzoticScaleLR}
a_{L,R} \left( t \right) = a_0 \left[ t_s - \left( \mp t \right)^{n} \right] ,
\ee
where the minus sign applies for the times $ t < 0 $ described by $a_L$ and the plus sign for the times $t>0$ described by $a_R$ (see Fig. \ref{scalefactor}).

%
%
In this scenario, the universe begins with a big-bang singularity at $ t_s = {\left(-t  \right)}^{n} $  for $- t_s  < t < 0$, faces an exotic singularity at $t=0$ (its type depends on the parameter $n$), then evolves towards a big-crunch singularity at $ t_s = t^{n} $  for $0  < t < t_s$ (one can also write a big-bang and a big-crunch times as $t = (\mp t_s)^{1/n}$).

The energy density and pressure for the scale factor (\ref{EgzoticScaleLR}) read as
\be
\rho_{L,R} = \frac{3}{8 \pi G} \left[ \frac{n^2 (\mp t)^{ 2 n -2}}{{\left[ t_s - \left( \mp t \right)^n \right]}^2} \right] ,
\label{rho1}
\ee
\be
p_{L,R} = - \frac{c^2}{8 \pi G} \left[ 2 \frac{n \left(1 - n \right) \left( \mp t \right)^{ n -2}}{\left[ t_s - \left( \mp t \right)^n \right]} + \frac{n^2 \left( \mp t \right)^{ 2 n -2}}{{\left[ t_s - \left( \mp t \right)^n \right]}^2} \right] .
\label{p1}
\ee
%
%
%
%
Using (\ref{rho1}) and (\ref{p1})  one can write the effective equation of state (though with an unseparable time) as follows
\be
p_{L,R} = - c^2 \left[ \frac{\rho}{3} \pm \frac{n-1}{\sqrt{6 \pi G}} \frac{\sqrt{\rho}}{t} \right].
\label{eos1}
\ee
From (\ref{eos1}) we immediately notice that in the limit $n=1$ we obtain the Friedmann universe with an equation of state for the cosmic strings fluid $p=-1/3 \rho$ \cite{AJ89} with the Penrose diagram covering the same region of the Einstein cylinder (\ref{ESU}) as the Minkowski metric (\ref{Mink}). This was presented in Fig. 3b of the Ref. \cite{harada}.

Now, we discuss two cases which are on both sides of the limit $n=1$: Finite Scale Factor Singularity (FSFS) and Sudden Future Singularity (SFS).

\subsection{Finite Scale Factor Singularity - FSFS}

For $ 0 < n < 1 $ we have FSFS at $t=0$ for the scale factor (\ref{EgzoticScaleLR}) which for
%
%
%
%
$ n = \frac{1}{2} $ leads to the conformal time (\ref{conftime}) given by
\be
\eta_{L,R} = \pm  \frac{2}{a_0} \left[ \sqrt{\mp t} - t_s \ln {(t_s - \sqrt{\mp t} )} \right] .
\label{SFSeta}
\ee
In more detail, we have for the left branch $ - {t_s}^2  \le t \le 0 $ that
\be
 - \infty \le \eta_L \le \frac{2 t_s \ln \left( t_s \right)}{a_0} = b ,
\ee
and for the right branch $ 0 \le t \le {t_s}^2 $ that
\be
 - b \equiv - \frac{2 t_s \ln \left( t_s \right)}{a_0} \le \eta_R \le \infty .
\ee
For the common time for both solutions $ t=0 $,  there is an FSFS with $ a(0) = a_0 t_s $, for $t = - t_s^2$ we have a big-bang singularity with $a=0$, while for $ t = {t_s}^2 $ we have a big-crunch singularity again with $ a = 0$.

\begin{figure}[htbp]
\includegraphics[width=8.3cm]{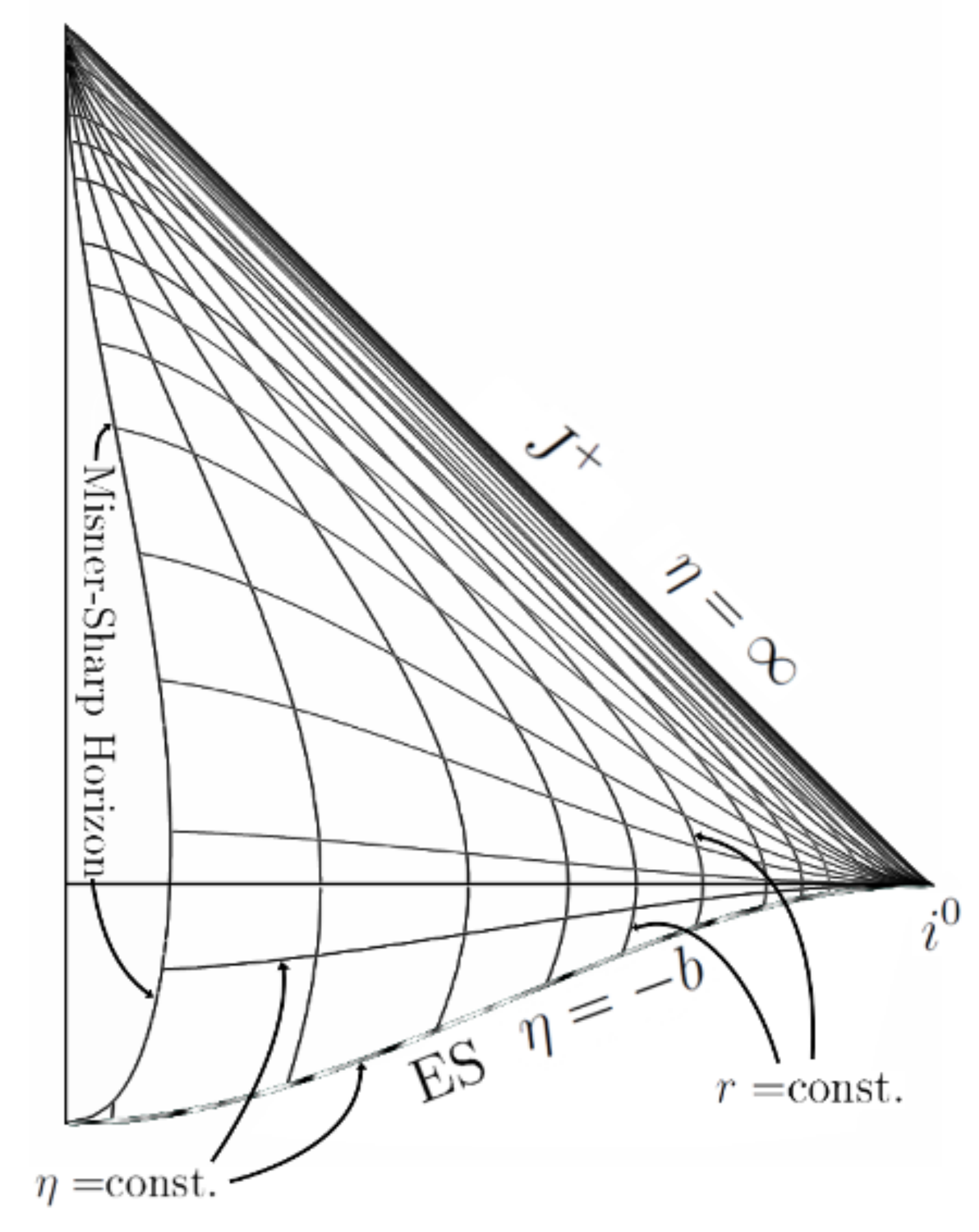}
\caption{Penrose Diagram for the right branch $a_R$ of the model (\ref{aLRB}) which begins with FSFS at the $t=0$ $(\eta = -b)$ hypersurface. Big-bang singularity $\eta = \infty)$ is isotropic $\cal{J}^+$. Misner-Sharp horizon (\ref{MScondition}) is on the left. ES is an exotic singularity.} 
\label{PenroseDiagrams1}
\end{figure}

\begin{figure}[htbp]
\includegraphics[width=8.3cm]{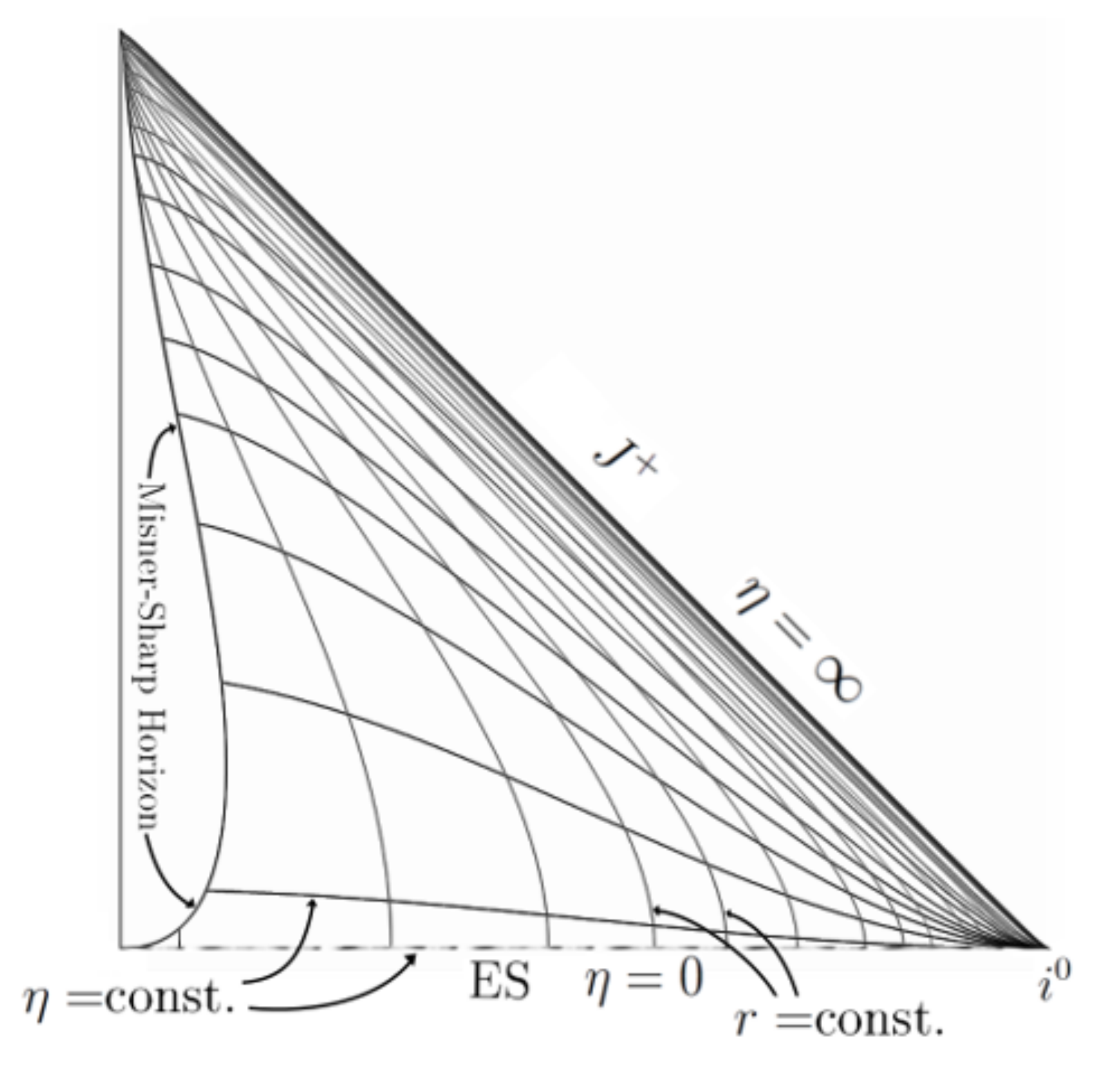}
\caption{Penrose Diagram for the right branch $a_R$ of the model (\ref{aLRB}) with FSFS which begins at the $t=0$ $(\eta = 0)$ hypersurface and for the parameter $t_s = 1$. Big-bang singularity $\eta = \infty)$ is isotropic $\cal{J}^+$. Misner-Sharp horizon (\ref{MScondition}) is on the left.}
\label{PenroseDiagrams2}
\end{figure}

\begin{figure}[htbp]
\includegraphics[width=8.3cm]{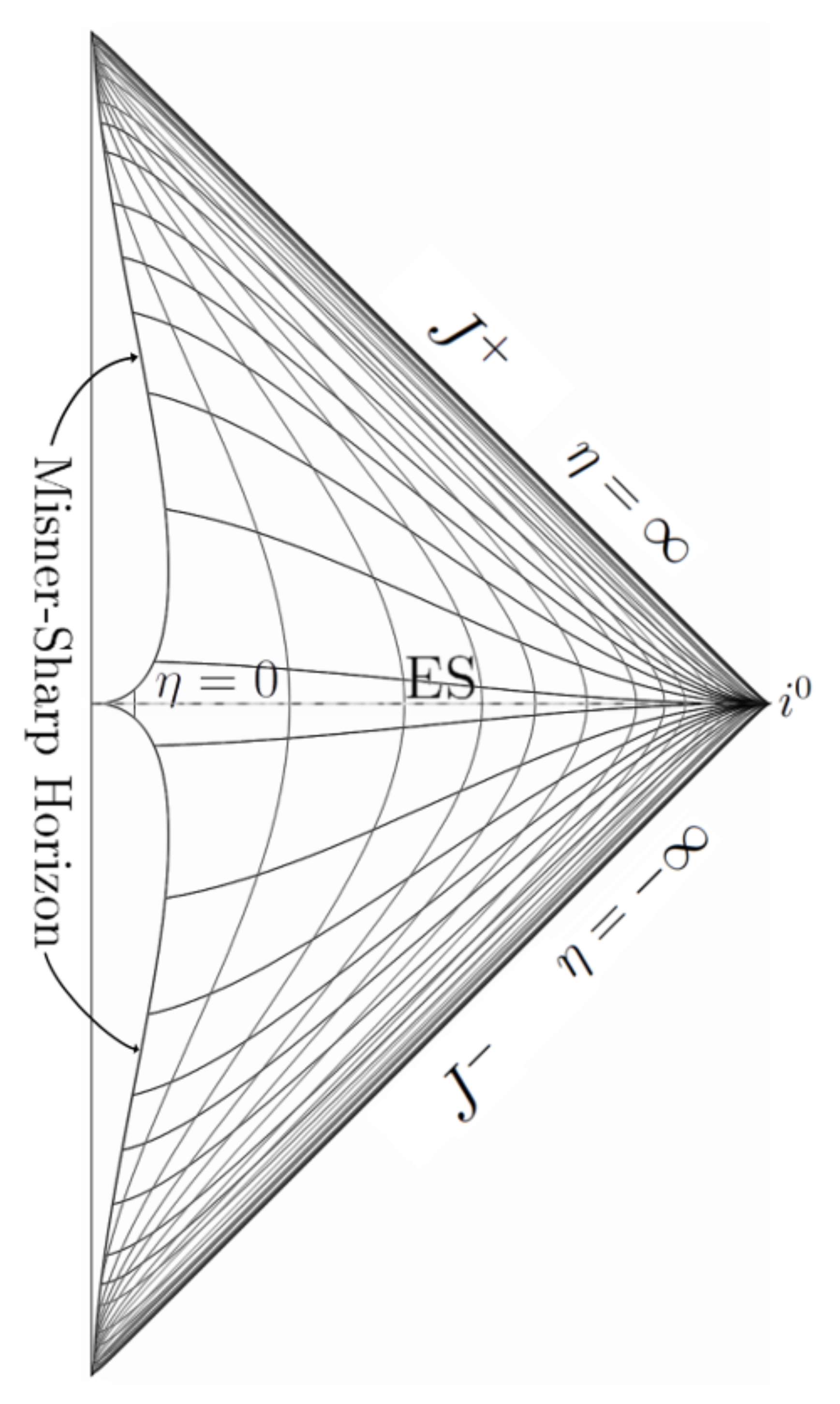}
\caption{Penrose Diagram for both branches of the model (\ref{aLRB}). It begins with a big-bang $(\eta = -\infty)$, then evolves to an FSFS at the $t=0$ $(\eta = 0)$ hypersurface ($t_s = 1$), and ends in big-crunch singularity ($\eta = \infty$). Both big-bang and big-crunch singularities are isotropic. Misner-Sharp horizons (\ref{MScondition}) are on the left.}
\label{PenroseDiagrams3}
\end{figure}

Analysing the ranges of the conformal time $\eta$ one can say that the first part of the left brach of our model (\ref{EgzoticScaleLR}) is mapped onto a piece of the Minkowski diagram with an initial big-bang singularity at $t=-t_s^2$, $\eta = - \infty$ which is isotropic and with a cut-off at the FSFS hypersurface $t=0$, $\eta = b >0$ which is spacelike. The right branch of (\ref{EgzoticScaleLR}), on the other hand, is mapped onto another piece of the Minkowski diagram starting with an FSFS hypersurface $t=0$, $\eta = -b >0$ which is spacelike, and then evolving towards the final big-crunch singularity at at $t=t_s^2$, $\eta = \infty$, which is isotropic (see Fig. \ref{PenroseDiagrams1}). The areas for left and right branches overlap in the region $-b < \eta < b$ and they have only one common hypersurface when $t_s = 1$. In such a case, the left branch is identical with a lower half of the Minkowski Penrose diagram, and the right branch is identical with an upper half of the Minkowski diagram as in Figs. \ref{PenroseDiagrams2} and \ref{PenroseDiagrams3}. 

One is able to invert the relation (\ref{SFSeta}) to get
\be
t(\eta) =\pm {t_s}^2 {\left[ 1 + W \left[ - \frac{1}{t_s} \exp \left(-1 - \frac{a_0 \eta}{2 t_s} \right) \right] \right]}^2 ,
\ee
where $W(z) = z \exp{(z)}$ is the Lambert function. Using the definition of an affine parameter one gets then
\bea
\lambda(t) &=&  \int a^2 \left( \eta \right) d \eta = \int a \left( t \right) dt , \\
&=& a_0 t \left[ t_s - \frac{2}{3} \left( \mp t \right)^{\frac{1}{2}} \right]  \nonumber ,
\eea
so that at singularities $\lambda(0) = 0$, and $\lambda(\pm t_s^2) = (1/3) a_0 t_s^3$, which means that the parameter is finite. It is also useful to calculate the Misner-Sharp mass \cite{MS,harada} which in our case gives
\be
\frac{2m}{a(t)} = \frac{a_0^2 r^2}{4t} ,
\ee
and so the past/future trapping regions are for 
\be
r > \pm \frac{2 t_s}{a_0} \left[ 1 + W \left[ - \frac{1}{t_s} \exp \left(-1 - \frac{a_0 \eta}{2 t_s} \right) \right] \right] .
\label{MScondition}
\ee

\subsection{Sudden Future Singularity - SFS}

For $ 1 < n < 2 $ we have an SFS. Let us take $ n = \frac{3}{2} $ as an example. In this case one gets the conformal time as 
\bea
\eta_{L,R} &=& \frac{1}{3 {a_0 t_s}^{\frac{1}{3}}} \left[ -2 \sqrt{3} \arctan \left( \frac{1 + 2 \sqrt{\mp t}}{\sqrt{3} {t_s}^{\frac{1}{3}}} \right)
\right.  \\
&-& \left. 2 \ln \left( {t_s}^{\frac{1}{3}}  - \sqrt{\mp t} \right) + \ln \left( {t_s}^{\frac{2}{3}} \mp t + \sqrt{\mp t} \right) \right] . \nonumber
\eea
For the right branch at $ t = 0 $ we have
\be
\eta = - \frac{\pi}{3 \sqrt{3} {a_0 t_s}^{\frac{1}{3}}} ,
\ee
and for $t = {t_s}^{\frac{2}{3}}$ we have $\eta = \infty$. For the left branch we have for $ t = 0 $ that
\be
\eta =  \frac{\pi}{3 \sqrt{3} {a_0 t_s}^{\frac{1}{3}}}
\ee
 and for $ t = - {t_s}^{\frac{2}{3}} $ we have  $ \eta = - \infty $.
 The parameter $ b $ is replaced onto $ - b $.

The Penrose diagram is similar as in the case of FSFS. 

\subsection{Hybrid big-rip/exotic singularity models}

\begin{figure}[htbp]
\includegraphics[width=8.3cm]{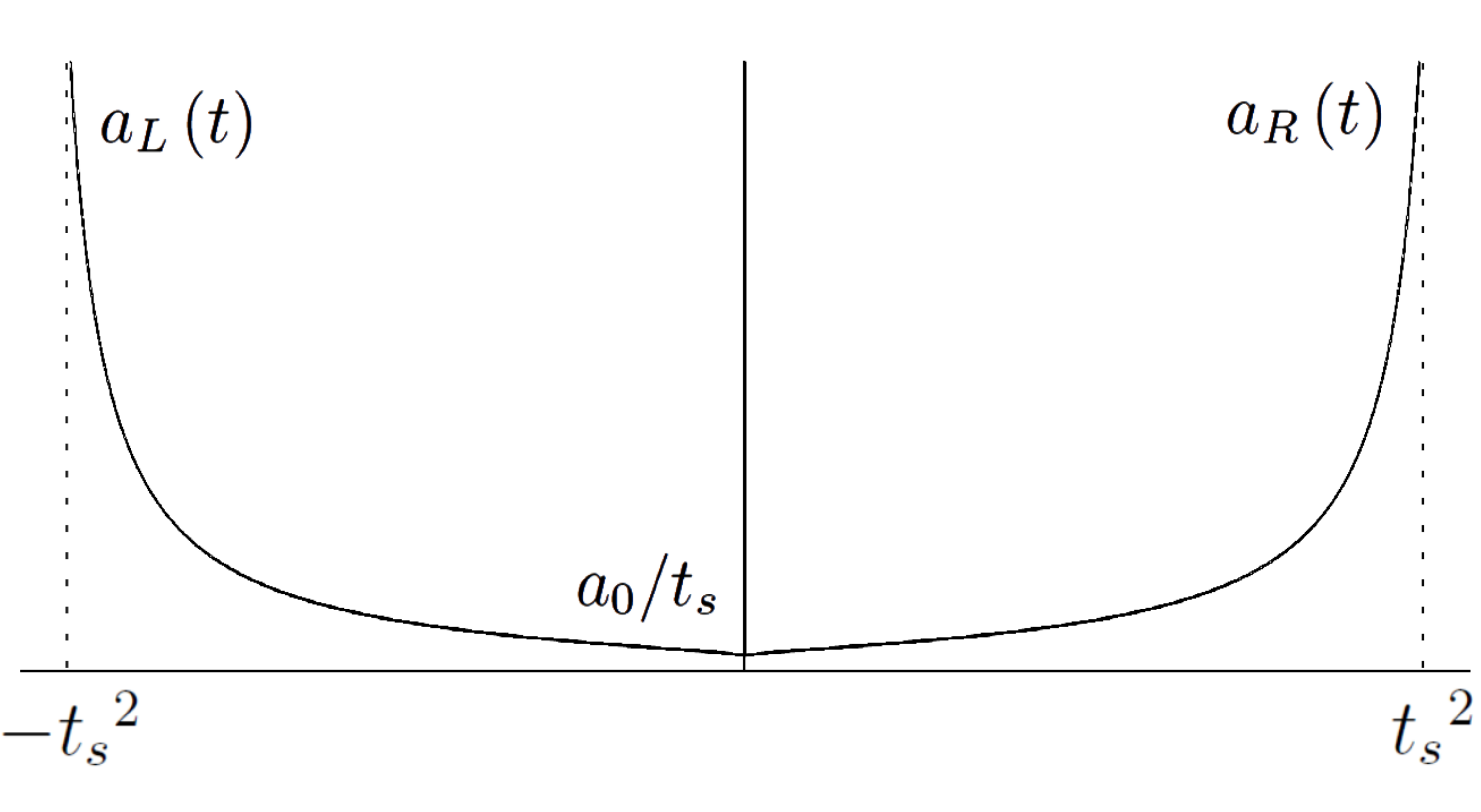}
\caption{The scale factor for the model (\ref{PSC}) with $ n = 1/2 $. The evolution starts with a big-rip, reaches an exotic singularity, and finally ends at another big-rip.}
\label{scalefactorBR}
\end{figure}

We can also select the model in the form 
\bea
\label{PSC}
a_{L,R} \left( t \right) = \frac{a_0}{t_s - \left(\mp t \right)^n} ~,
\eea
which starts at the big-rip for $ t = - {t_s}^{1/n} $, continues to an exotic singularity at $ t \to 0 $, and ends at another big-rip (anti-big-rip) at  $ t =  {t_s}^{1/n} $ as in Fig. \ref{scalefactorBR}. The density and the pressure functions are as follows:
\bea
\rho_{L,R} \left( t \right) &=& \frac{3}{8 \pi G} \left[ \frac{n^2 {\left(\mp t \right)}^{2n-2}}{{t_s-\left( \mp t \right)}^2}   \right]~,\\
p_{L,R} \left( t \right) &=& \frac{n c^2}{8 \pi G} \left[ \frac{\left( 2 + 3 n \right) {\left( \mp t \right)}^{2n -2} + \left( n - 1 \right) {\left( \mp t \right)}^{n-2} t_s}{{t_s-\left( \mp t \right)}^2} \right]~\nonumber \\ .
\eea
\begin{figure}[htbp]
\includegraphics[width=8.0 cm]{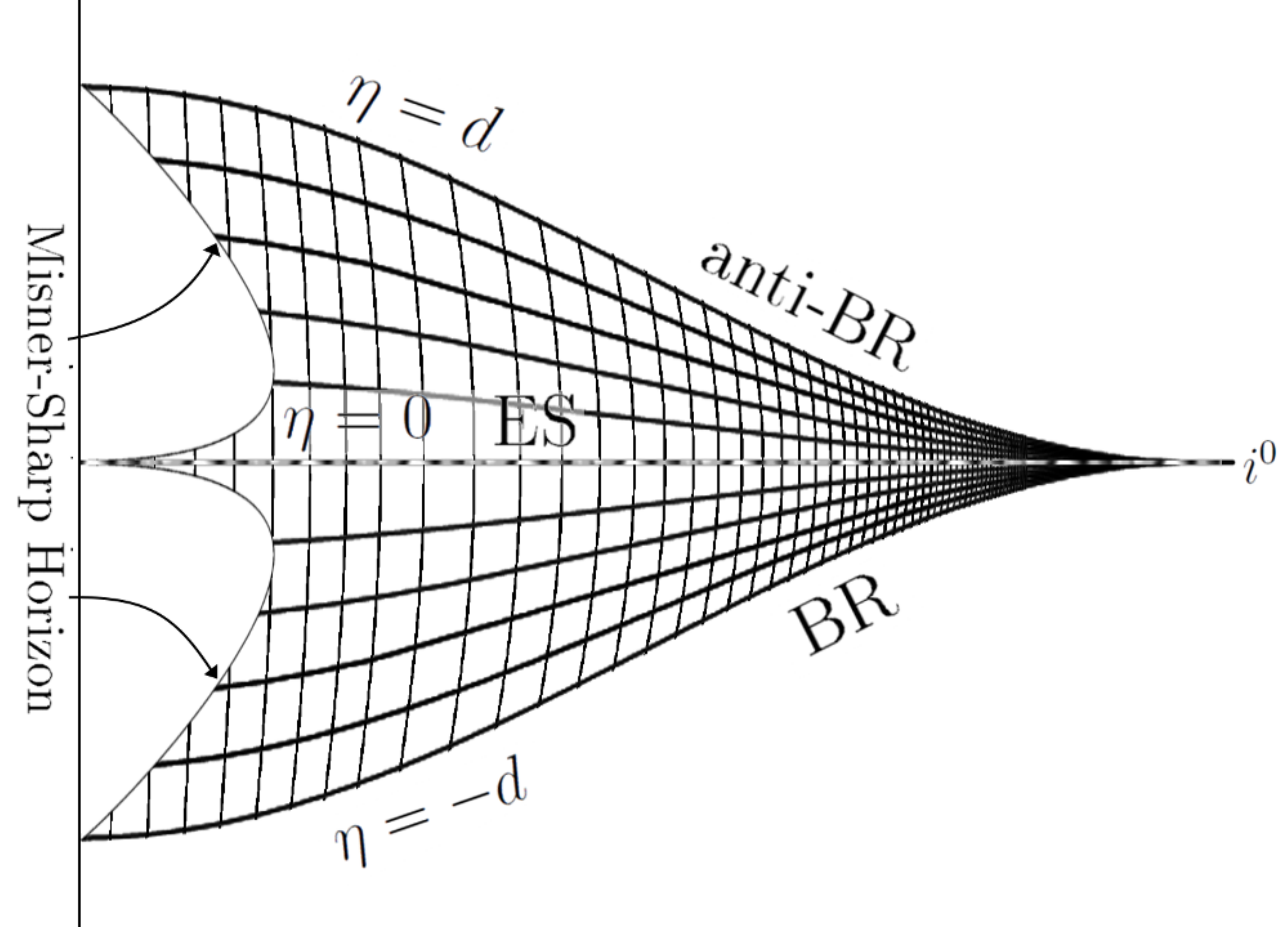}
\caption{Penrose diagram for the model (\ref{PSC}) with $ a_0=1 $, $t_s = 1 $ and $ n = 1/2 $ . The evolution begins with a big-rip singularity on a constant time hypersurface $t=-t_s^2$, evolves to an exotic singularity (here FSFS) to finally reach another big-rip (anti-big-rip) on a constant time hypersurface $t=t_s^2$. The Misner-Sharp horizons are on the left.}
\label{PenroseDiagrams4}
\end{figure}

The effective equation of state takes the form:
\bea
p_{L,R}= - \frac{c^2 \rho}{3n} \left[ 3n+2 + \left( 2n-2 \right) \frac{t_s}{{\left( \mp t \right)}^n} \right]
\eea
Conformal time for (\ref{PSC}) is:
\bea
\eta_{L,R}(t) = \frac{t_s}{a_0} \left[ t \pm \frac{ {\left( \mp t \right)}^{n+1}  }{\left( 1+n \right) t_s} \right] ~.
\label{t_eta_BR}
\eea
For $ t = 0 $ the conformal time $ \eta = 0 $ (exotic singularity), while for $t=(\mp t_s)^{1/n}$, $\eta = \mp n t_s^{\left( 1+n \right)/n}/ \left[ a_0 \left( 1+n \right) \right] $ (big-rip). The formula (\ref{t_eta_BR}) can be inverted for $n=1/2$ as follows
\bea
t \left( \eta \right) &=& \frac{1}{8} \left[ \mp 6 {t_s}^2 \mp \frac{ 3^{\frac{2}{3}} \left( 1 - i \sqrt{3} \right) \left( 8 a_0 t_s \eta \pm 3 {t_s}^4 \right)}{ X^{\frac{1}{3}}} \right. \nonumber \\
&-& \left. 3^{\frac{1}{3}} \left( 1 + i \sqrt{3} \right) X^{\frac{1}{3}} \right]
\eea
where
\bea
X &=& \mp 9 {t_s}^6 - 36 a_0 {t_s}^3 \eta \mp 24 {a_0}^2 {\eta}^2\\
&+&  8 \sqrt{ 3 {a_0}^3 {\eta}^3 \left( 3 a_0 \eta \pm {t_s}^3 \right)} \nonumber ~.
\eea
In this case for $ \eta = 0 $ we have $ t = 0 $, while for  $\eta = \mp t_s^3/3a_0 \equiv \mp d$, we have $t=(\mp t_s)^{2}$. 
In the $ n = 1/2 $ model which corresponds to an FSFS we have the affine parameter 
\bea
\lambda(t) &=&  \int a^2 \left( \eta \right) d \eta = \int a \left( t \right) dt , \\
&=& 2a_0 \left( t_s - \sqrt{t} - t_s \ln{\mid t_s - \sqrt{t} \mid } \right)  \nonumber ,
\eea
so that at an exotic singularity $\lambda(0) = 2a_0 t_s (1 - \ln{t_s})$, and at big-rip $\lambda(\mp t_s^2) = \mp \infty$, which proves geodesic incompleteness of the latter. 

The Misner-Sharp mass reads as
\bea
\frac{2m}{a(t)}= \frac{ {a_0}^2 n^2 r^2 {\left( \mp t \right)}^{2n-2}}{{\left[ t_s - \left( \mp t \right) \right]}^4}
\eea
and the condition for the trapping horizon is 
\bea
r > \pm \frac{ {\left( \pm t\right)}^{1-n} { \left[ { \left( \pm t \right)}^n - t_s \right]}^2 }{a_0 n}~.
\eea
The Penrose diagram for the model (\ref{PSC}) with $ a_0=1 $, $t_s = 1 $ and $ n = 1/2 $ is plotted in Fig. \ref{PenroseDiagrams4}. The evolution begins with a big-rip singularity on a constant time hypersurface $t=-t_s^2$, evolves to an exotic singularity (here FSFS) to finally reach another big-rip (which we call an anti-big-rip in order to make a difference with an ''initial" big-rip in full analogy to big-bang/big-crunch differentiation).

\subsection{Big-separation and $w$-singularity‡}

For $2 < n < 3 $ in (\ref{EgzoticScaleLR}) one obtains a big-separation (BS) singularity, while for $ 3 < n < 4 $ a $w$-singularity. The conformal diagrams are analogous. There is a duality between the SFS, BS, and $w$-singularity models and FSFS, and big-rips models and the dividing line is $n=1$. It can be considered ''phantom duality'' type symmetry \cite{phantom}. 

\subsection{Classical analogues} 

There is a nice analogy between an SFS singularity of pressure and an acceleration singularity at the start of a car at car-drag races (cf. footnote 1 of Ref. \cite{cotsakis}). We may extend such an analogy into other exotic singularities assuming that the  power scales as follows (in Ref. \cite{cotsakis} only the case $p=0$ is considered):
\be
P = v \dot{v} \sim t^p ,
\ee
where $v$ is the velocity, $\dot{v} = a$ is the acceleration, and $p=$ const. 
After integrating we have 
\be
v \propto t^{\frac{1}{2}(s+1)}, \hspace{0.5cm}  a \propto t^{\frac{1}{2}(s-1)} .
\label{va}
\ee
We can also define a derivatives of acceleration (being like jerk and snap in cosmology \cite{kamionk}) as 
\be
\dot{a} = \propto t^{\frac{1}{2}(s-3)}, \hspace{0.5cm}  \ddot{a} \propto t^{\frac{1}{2}(s-5)} .
\label{aaaa}
\ee
The following conclusions can be deducted from (\ref{va}) and (\ref{aaaa}). The velocity is singular provided $s<-1$, while the acceleration is singular provided $v<1$ etc. The former case is an analogue of a density singularity in general relativity, while the latter is an analogue of a pressure singularity etc. In conclusion one can say that a classical analogue of an FSFS singularity is when $s<-1$ (which corresponds to $1<n<2$ in (\ref{aLRB})), of an SFS singularity when $-1<s<1$ ($0<n<1$), of a BS singularity when $1<s<3$ ($2<n<3$), and of a $w$-singularity when $3<s<5$ ($3<n<4$). 

\subsection{Raychaudhuri averaging}

Following \cite{PLB11} we can calculate the acceleration scalar for any Friedmann model as 
\be
\chi = \theta_{,\mu} u^{\mu}
= 3 H^2 \left( q + 1 \right)
\ee
where $\theta$ is the expansion scalar, $u^{\mu}$ the four-velocity of a comoving observer, $ H= \dot{a} / a $ the Hubble parameter and $ q = - \ddot{a} a / {\dot{a}}^2 $ the deceleration parameter.

Then, we may calculate the so-called Raychaudhuri averaging \cite{Raych} of such an average acceleration scalar in flat Friedmann background to get
\be
< \dot{\theta} > = \frac{3 \int_{t_0}^{t_1} \frac{{\dot{a}}^2}{a^2} \left( - \frac{\ddot{a} a}{{\dot{a}}^2} +1 \right) dt}{\int_{t_0}^{t_1} a^3 dt}
\ee
For scale factor (\ref{aLRB}) and $ n= 3/2 $ which corresponds to an SFS we obtain (for both branches $ a_L $ and $ a_R $)
\be
< \dot{\theta} > = \frac{3 \int_{0}^{{t_s}^2} \frac{{\dot{a}}^2}{a^2} \left( - \frac{\ddot{a} a}{{\dot{a}}^2} +1 \right) dt}{\int_{0}^{{t_s}^2} a^3 dt}=-\frac{165}{14 {t_s}^{\frac{4}{3}}} = const. ,
\ee
which means that this average is finite. The same is true for a BS and a $w$-singularity while for an FSFS $ < \dot{\theta} > $ blows up to infinity and so it can be considered as "strong" singularity in view of Raychaudhuri averaging. The same is the case for a big-rip singularity (given for example by the scale factor (\ref{PSC}) which is then stronger in the sense of Raychaudhuri averaging than a big-bang for which the average is finite \cite{PLB11}. 

\section{Conclusion - beyond singularities}
\label{beyond}

We have investigated the conformal structure of exotic singularity universes and presented their appropriate Penrose diagrams. We have found that the conformal structure of these exotic singularities is not very much ''exotic'' since they are just constants time hypersurfaces in the diagrams.  

Our discussion of the Penrose diagrams of the weak exotic singularities suggests that they are transversable since there is no geodesic incompleteness. This happens despite there are discontinuities of energy density, pressure, derivatives of pressure, or other physical quantities \cite{adam,yurov} not only in homogeneous configurations, but also in anisotropic and inhomogeneous backgrounds \cite{cotsakis}. By using the method of Penrose diagrams one is able to study the transversability quite systematically. Our method relies on the fact of geodesic completeness and nicely allows to present both (left -- before singularity, and right -- after singularity) phases of the evolution of the universes which possesses both weak (e.g. SFS) and strong (e.g. big-rip) singularities. One of a possible application could be gluing hybrid big-bang to weak exotic singularity (half-)diagram with a weak exotic singularity to an inhomogeneous big-bang singularity model which allows spatial pressure singularities \cite{JMP93}. Such a possibility would allow a timelike singularity of pressure (an SFS) being converted at a transition into a Finite Density singularity of pressure which is present in some spatial regions of the universe through the whole second piece of the hybrid evolution. In fact, in such a model there would be an interesting effect of a "leakage" of pressure infinity through some "spatial holes" only while transiting a weak singularity and then evolving towards another big-bang. The detailed studies of such exotic options in less symmetric geometries will be the matter of future studies.

%
%

\begin{acknowledgements}
This project was financed by the Polish National Science Center Grant DEC-2012/06/A/ST2/00395.
\end{acknowledgements}


\begin{thebibliography}{}
\bibitem{HE} S.W. Hawking and G.F.R. Ellis, {\it The large-scale structure of space-time} (Cambridge Univ. Press, Cambridge, 1999)~.

\bibitem{tipler} F.J. Tipler, Phys. Lett. A{\bf 64}, 8 (1977).

\bibitem{krolak} A. Kr\'olak, Classical Quantum Grav. {\bf 3}, 267 (1988).

\bibitem{supernovae} S. Perlmutter et al., Astroph. J. \textbf{517}, (1999) 565; A. G. Riess et al., Astron. J. \textbf{116}, 1009 (1998); A.G. Riess et al., Astroph. J. \textbf{560}, 49 (2001), J.L. Tonry et al., Astroph. J. \textbf{594}, 1 (2003); M. Tegmark et al., Phys. Rev. \textbf{D69}, 103501 (2004); R.A. Knop et al., Astrophys. J. \textbf{598}, 102 (2003), M. Kowalski et al., Astrophys. J. \textbf{686}, 749 (2008).

\bibitem{phantom} R.R. Caldwell, Phys. Lett. B {\bf 545}, 23 (2002); B. MacInnes, JHEP {\bf 0208}, 029 (2002); R.R. Caldwell, M. Kamionkowski, and N.N. Weinberg, Phys. Rev. Lett. {\bf 91}, 071301 (2003); S.M. Carroll, M. Hoffman, and A. Trodden, Phys. Rev. D{\bf 68}, 023509 (2003); M.P. D\c{a}browski, T. Stachowiak and M. Szyd{\l }owski, Phys. Rev. D {\bf 68}, 103519 (2003); S. Nojiri and S.D. Odintsov, Phys. Rev. D{\bf 70}, 103522 (2004); P. F. Gonz\'alez-D\'iaz, Phys. Rev. D{\bf 68}, 021303; Phys. Lett. B{\bf 586}, 1 (2004); M. Bouhmadi-Lopez and J. A. Jim\'enez-Madrid, Journ. Cosmol. Astrop. Phys. {\bf 0505}, 005 (2005); M. Bouhmadi-Lopez and P. F. Gonz\'alez-D\'iaz, Phys. Lett. B{\bf 659}, 1 (2008); K. Bamba, S. Nojiri, and S.D. Odintsov, Journ. Cosmol. Astrop. Phys. {\bf 0810}, 045 (2008). 

\bibitem{Barrow04} J.D. Barrow, Class. Quantum Grav. {\bf 21}, L79 (2004). 

\bibitem{big-brake} V. Gorini, A. Kamenshchik, U. Moschella, and V. Pasquier, Phys. Rev. D{\bf 69}, 123512 (2004).

\bibitem{obsSFS} M.P. D\c{a}browski, T. Denkiewicz, M.A. Hendry, Phys. Rev. D{\bf 75}, 123524 (2007); Z. Keresztes, L.A. Gergely, V. Gorini, U. Moschella, A.Yu. Kamenshchik,
Phys. Rev. D{\bf 79}, 083504 (2009); Z. Keresztes, L.A. Gergely, A.Yu. Kamenshchik, V. Gorini, D. Polarski, Phys. Rev. D{\bf 82}, 123534 (2010); H. Ghodsi, M.A. Hendry, M.P. D\c{a}browski, T. Denkiewicz, Mon. Not. Roy. Astron. Soc. (2011).

\bibitem{exotic} J.D. Barrow, Class. Quantum Grav. {\bf 21}, 5619 (2004); K. Lake, Class. Quantum Grav. {\bf 21}, L129 (2004); S. Nojiri and S.D. Odintsov, Phys. Lett. B{\bf 585}, 1 (2004); K. Bamba, S.D. Odintsov, L. Sebastiani, and S. Zebrini, Eur. Phys. Journ. C{\bf 67}, 295 (2010); A.V. Astashenok, S. Nojiri, S.D. Odintsov, and R.J. Scherrer, Phys. Lett. B{\bf 713}, 145 (2012); J. Beltr\'an Jim\'enez, D. Rubiera-Garcia, D. S\'aez-G\'omez, and V. Salzano, Phys. Rev. D{\bf 94}, 123520 (2016). 

\bibitem{nojiri} S. Nojiri, S.D. Odintsov and S. Tsujikawa, Phys. Rev. D {\bf 71},063004 (2005).

\bibitem{wsing} M. P. D\c{a}browski and T. Denkiewicz, Phys. Rev. D {\bf 79}, 063521 (2009); L. Fernandez-Jambrina, Phys. Rev. D{\bf 82}, 124004 (2010).

\bibitem{LRip} P.H. Frampton, K.J. Ludwick, R.J. Scherrer, Phys. Rev. D{\bf 84}, 063003 (2011).

\bibitem{PRip} P.H. Frampton, K.J. Ludwick, R.J. Scherrer, Phys. Rev. D{\bf 85}, 083001 (2012).

\bibitem{APS2010} M.P. D\c{a}browski and T. Denkiewicz, AIP Conference Proceedings {\bf 1241}, 561 (2010); arXiv: 0910.0023.

\bibitem{limits} M.P. D\c{a}browski, {\it Mathematical Structures of the Universe}, M. Eckstein, M. Heller, S.J. Szybka (eds.) (Copernicus Center Press, Krak\'ow, 2014) p. 99; arXiv: 1407.4851.

\bibitem{PRD05} M.P. D\c{a}browski, Phys. Rev. D{\bf 71}, 103505 (2005). 


\bibitem{adam} M.P. D\c{a}browski and A. Balcerzak, Phys. Rev. D{\bf 73}, 101301(R) (2006).

\bibitem{yurov} A.V. Yurov, A.V. Astashenok, V.A. Yurov, Eur. Phys. Journ. C{\bf 78}, 542 (2018).

\bibitem{cotsakis} J.D. Barrow and S. Cotsakis, Phys. Rev. D{\bf 88}, 067301 (2013).  

\bibitem{LFJ2007} L. Fernandez-Jambrina, Phys. Lett. B{\bf 656}, 9 (2007).

\bibitem{leonardo1} L. Fernandez-Jambrina and R. Lazkoz, Phys. Rev.
D{\bf 70}, 121503(R) (2004); L. Fernandez-Jambrina and R. Lazkoz, Phys. Rev. D{\bf 74}, 064030 (2006).

\bibitem{perivolaropoulos} L. Perivolaropoulos, Phys. Rev. D{\bf 94}, 124018 (2018). 

\bibitem{kamen17} A. Kamenshchik, Found. Phys. {\bf 48}, 1159 (2018). 

\bibitem{brandenb18} R. Brandenberger, Phys. Rev. D{\bf 98}, 063521 (2018).  

\bibitem{harada} T. Harada, B.J. Carr, and T. Igata, Class. Quantum Grav. {\bf 35}, 105011 (2018). 

\bibitem{PLB11} M. P. D\c{a}browski, Phys. Lett. B{\bf 702}, 320 (2011).

\bibitem{AJ89} M.P. D\c{a}browski and J. Stelmach, Astron. Journ. {\bf 97}, 978 (1989).

\bibitem{MS} C.W. Misner and C.H. Sharp, Phys. Rev. {\bf 136}, B571 (1964). 




\bibitem{JCAP13} M.P. D\c{a}browski and K. Marosek, Journ. Cosmol. Astrop. Phys. {\bf 02}, 012 (2013).

\bibitem{plb05} M.P. D\c{a}browski, Phys. Lett. B {\bf 625}, 184 (2005).

\bibitem{kamionk} R.R. Caldwell and M. Kamionkowski, Journ. Cosmol. Astrop. Phys. {\bf 0409}, 009 (2004).

\bibitem{Raych} A.K. Raychaudhuri, Phys. Rev. Lett. {\bf 80}, 654 (1998).

\bibitem{JMP93} M.P. D\c{a}browski, Journ. Math. Phys. {\bf 34}, 1447 (1993). 

%
%
\end{thebibliography}


\end{document}